\documentclass[12pt,a4paper]{article}
\usepackage{amsmath,amssymb}
\usepackage[sort]{cite}
\usepackage{graphicx}
\def\mmu{m_{\mu}}
\def\iep{i\varepsilon}
\def\arctanh{{\rm arctanh}}
\newcommand{\Li}[1]{{\rm Li}_{#1}}
\newcommand{\amu}[1]{a^{\text{HVP}(#1)}_{\mu}}
\newcommand{\APF}[1]{A_{0}^{(#1)}}
\newcommand{\KG}[2]{K_{#1}^{(#2)}}
\newcommand{\KGt}[2]{\tilde{K}_{#1}^{(#2)}}

\begin{document}

\begin{center}

{\Large\bf Addendum: Timelike and spacelike kernel functions for \\
the hadronic vacuum polarization contribution \\
to the muon anomalous magnetic moment 

(2022 J.\ Phys.\ G: Nucl.\ Part.\ Phys.~\textbf{49} 055001)

}

\vskip10mm

{\large A.V.~Nesterenko}

\vskip7.5mm

{\small\it Bogoliubov Laboratory of Theoretical Physics,
Joint Institute for Nuclear Research,\\
Dubna, 141980, Russian Federation}

\end{center}

\vskip5mm

\noindent
\centerline{\bf Abstract}

\vskip2.5mm

\centerline{\parbox[t]{150mm}{%
This addendum provides results complementary to those obtained in
[J.~Phys.~G~\textbf{49}, 055001 (2022)]. Specifically, an~equivalent 
form of the relation, which binds together the ``spacelike'' kernel 
functions for the hadronic vacuum polarization contribution to the 
muon anomalous magnetic moment~$a^{\text{HVP}}_{\mu}$, is~obtained. 
It~is shown that the infrared limiting value of the ``spacelike'' 
and ``timelike'' kernel functions, which enter the representations 
for~$a^{\text{HVP}}_{\mu}$ involving the Adler function and the 
\mbox{$R$--ratio}, is~identical to the corresponding QED~contribution 
to the muon anomalous magnetic moment of the preceding order in the 
electromagnetic coupling. The~next--to--leading order 
contributions~$\amu{3b}$ (which includes the leptonic and hadronic 
insertions) and~$\amu{3c}$ (which includes the double hadronic 
insertion), are~studied. The~three kernel functions appearing in 
the representations for~$\amu{3b}$, which involve the hadronic 
vacuum polarization function, Adler function, and the $R$--ratio, 
are presented for the cases of the electron and \mbox{$\tau$--lepton} 
loop insertions.
\\[2.5mm]
}}

\vskip12mm

\setcounter{equation}{47}
\setcounter{figure}{6}

This~addendum provides results complementary to those obtained in Ref.~\cite{JPG49}.
In~what follows the notations as well as the numbered references to equations, 
figures, sections, and bibliography correspond to those of Ref.~\cite{JPG49} unless 
otherwise explicitly specified. The~enumeration of equations, figures, and references 
hereinafter continues that of Ref.~\cite{JPG49}.

\bigskip

\begin{figure}[t]
\centerline{\includegraphics[width=75mm,clip]{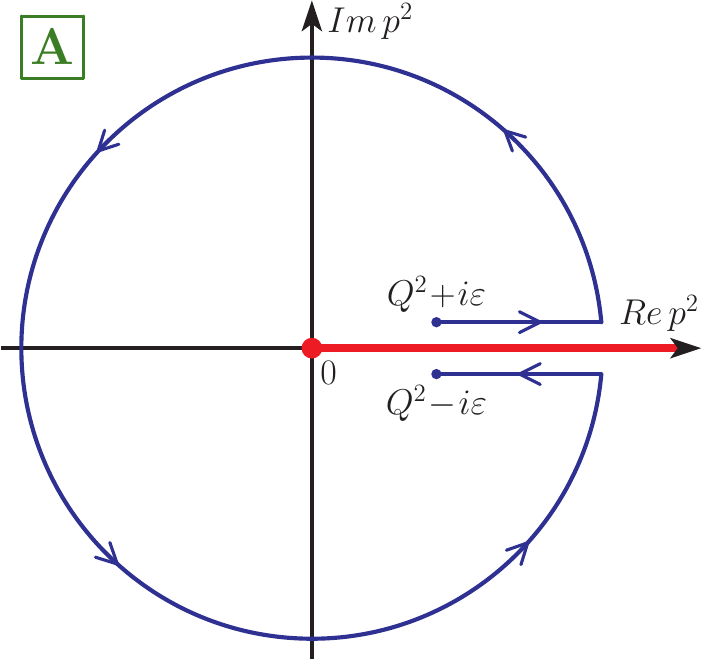}%
\hspace{10mm}%
\includegraphics[width=75mm,clip]{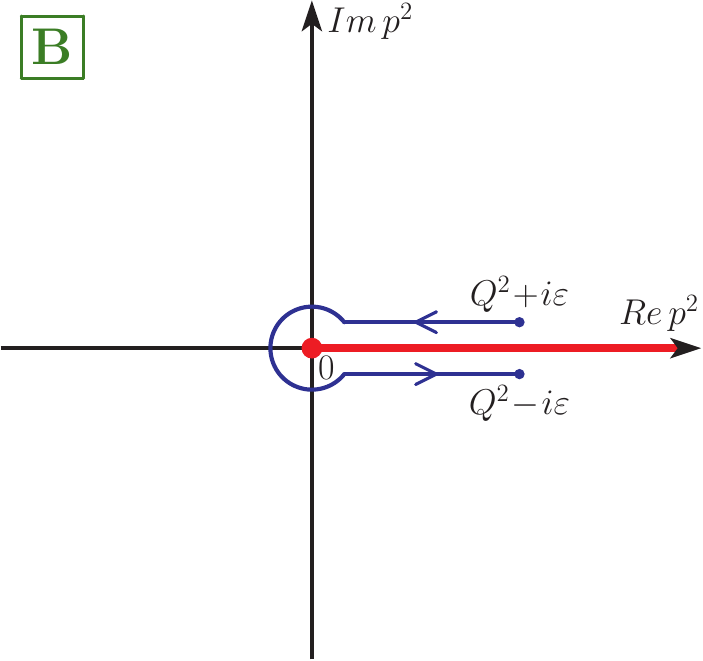}}
\caption{The~integration contours in the complex~$p^2$--plane
in Eqs.~(33),~(\ref{KRelDR}). The~physical cut $p^2 \ge 0$ of 
the ``timelike'' kernel function~$K_{R}(-p^2)$~(15c) is shown 
along the positive semiaxis of real~$p^2$.}
\label{Plot:Contour2}
\end{figure}

First of all, Eq.~(31) implies that the relation~(30), which expresses the kernel 
function~$K_{D}(Q^2)$ in terms of~$K_{\Pi}(Q^2)$, can also be represented in an 
equivalent form, which is particularly convenient for use at low energies:
\begin{equation}
\label{KRelDP}
K_{D}(Q^2) = \frac{1}{Q^2} 
\int\limits_{Q^2}^{\infty}\!\! K_{\Pi}(\xi)\, d \xi
=
\frac{4\mmu^{2}}{Q^2}K_{0} - 
\frac{1}{Q^2} 
\int\limits_{0}^{Q^2}\!\! K_{\Pi}(\xi)\, d \xi.
\end{equation}
In~this equation~$\xi = -p^2 \ge 0$ stands for a spacelike kinematic variable of the dimension 
of~$\mbox{GeV}^2$ and $K_{0}$~denotes the infrared limiting value of the respective ``spacelike''
and ``timelike'' functions, namely
\begin{equation}
\label{KRDlim}
K_{0} = 
\lim_{Q^2 \to 0_{+}} \frac{Q^2}{4\mmu^{2}}\, K_{D}(Q^2) =
\lim_{s \to 0_{+}} \frac{s}{4\mmu^{2}}\, K_{R}(s) =
\int\limits_{0}^{\infty}\!\! K_{\Pi}(\xi)\, \frac{d \xi}{4\mmu^{2}} .
\end{equation}
The first equality in Eq.~(\ref{KRelDP}) corresponds to the use
in relation~(33)
\begin{equation}
\label{KRelDR}
K_{D}(Q^2) = - \frac{1}{2 \pi i} \lim_{\varepsilon \to 0_{+}}
\frac{1}{Q^2} 
\int\limits_{Q^2 + \iep}^{Q^2 - \iep}
K_{R}(-p^2) d p^2
\end{equation}
the integration contour displayed in Fig.~\ref{Plot:Contour2}$\,$A (here 
the integration along the circle of infinitely large radius vanishes). 
In~turn, the second equality in Eq.~(\ref{KRelDP}) corresponds to the 
use in~Eqs.~(33) and~(\ref{KRelDR}) the integration contour shown in 
Fig.~\ref{Plot:Contour2}$\,$B. In~this case the integration along the 
circle of vanishing radius yields the term proportional to~$K_0$
on the right--hand side of~Eq.~(\ref{KRelDP}).

It~is necessary to outline that the last equality in Eqs.~(31) and~(\ref{KRDlim}) 
implies that the infrared limiting value~$K_{0}$ is identical to the corresponding 
QED~contribution to the muon anomalous magnetic moment of the preceding order in 
the electromagnetic coupling~$\alpha$. In~particular, in~the second order in~$\alpha$ 
Eq.~(\ref{KRDlim}) takes the following form
\begin{equation}
\label{Klim2}
K_{0}^{(2)} = 
\lim_{Q^2 \to 0_{+}} \frac{Q^2}{4\mmu^{2}} \KG{D}{2}(Q^2) =
\lim_{s \to 0_{+}} \frac{s}{4\mmu^{2}}\KG{R}{2}(s) =
\int\limits_{0}^{\infty}\!\! \KG{\Pi}{2}(\xi)\, \frac{d \xi}{4\mmu^{2}}  = \frac{1}{2}\,  ,
\end{equation}
that constitutes the leading Schwinger contribution~\cite{JS48}, see Eqs.~(10), (16), 
(34), (35), and~Fig.~5. 

\bigskip

As~mentioned in Sect.~2.2, in the next--to--leading order of perturbation theory
(i.e.,~in the third order in the electromagnetic coupling) the hadronic vacuum 
polarization contribution to the muon anomalous magnetic moment consists of 
three parts, namely
\begin{equation}
\label{Amu3Def}
\amu{3} = \amu{3a} + \amu{3b} + \amu{3c}.
\end{equation}
The first term on the right--hand side of this equation (see~Fig.~2) has been 
addressed in~Sect.~3.2 and in~this case Eq.~(\ref{KRDlim}) reads~(39)
\begin{align}
\label{Klim3a}
K_{0}^{(3a)} & = 
\lim_{Q^2 \to 0_{+}} \frac{Q^2}{4\mmu^{2}}\, \KG{D}{3a}(Q^2) =
\lim_{s \to 0_{+}} \frac{s}{4\mmu^{2}}\, \KG{R}{3a}(s) =
\int\limits_{0}^{\infty}\!\! \KG{\Pi}{3a}(\xi)\, \frac{d \xi}{4\mmu^{2}}  = 
\nonumber \\ & =
\frac{197}{144} +\frac{1}{2}\zeta_{2} 
-3\zeta_{2}\ln(2) +\frac{3}{4}\zeta_{3} 
\simeq -0.328479,
\end{align}
where $\zeta_{t}$ stands for the Riemann $\zeta$~function~(40). Equation~(\ref{Klim3a}) 
constitutes the QED~contribution obtained in Refs.~\cite{Sommerfield1957, Sommerfield1958, 
Petermann1957, Petermann1958}, see Eqs.~(36), (30), (43), and Fig.~6.

\begin{figure}[t]
\centerline{%
\begin{tabular}{lcr}
\includegraphics[height=50mm,clip]{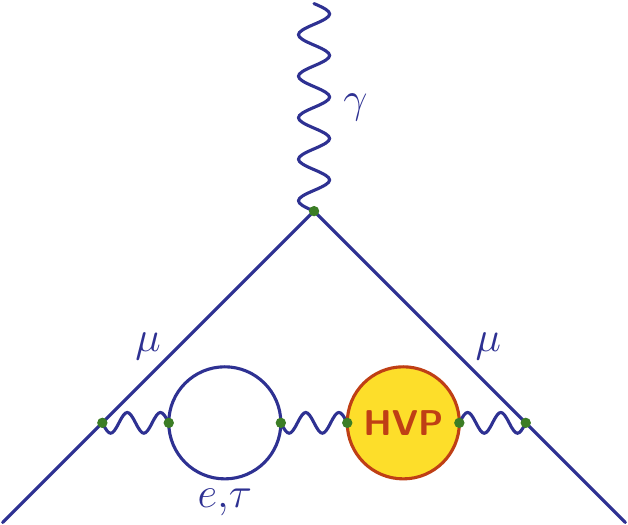} & &
\includegraphics[height=50mm,clip]{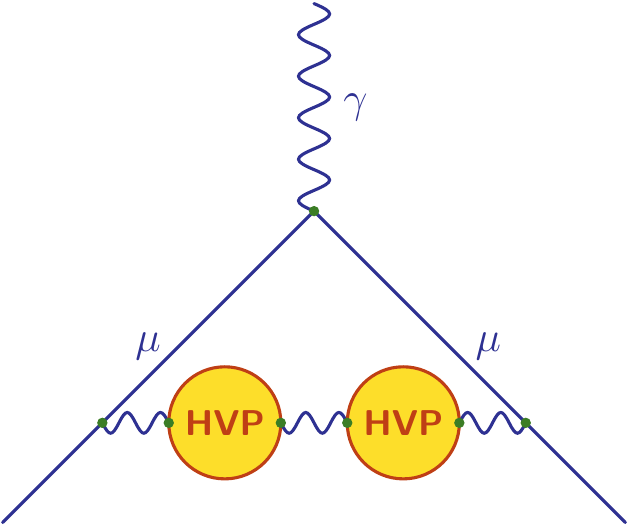} \\ 
\end{tabular}}
\caption{The diagrams contributing to the second [$\amu{3b}$, left plot] 
and third [$\amu{3c}$, right plot] terms of the next--to--leading order 
hadronic vacuum polarization contribution to~the muon anomalous magnetic 
moment~(\ref{Amu3Def}).}
\label{Plot:Amu3bc}
\end{figure}

In~turn, the second term on the right--hand side of Eq.~(\ref{Amu3Def}) corresponds to 
the diagrams, which include one closed electron (or~\mbox{$\tau$--lepton}) loop (see left 
plot of Fig.~\ref{Plot:Amu3bc}), specifically
\begin{equation}
\label{Amu3bDef}
\amu{3b} = \frac{2}{3} \Bigl(\frac{\alpha}{\pi}\Bigr)^{\!3}
\!\int\limits_{s_{0}}^{\infty}
\frac{G_{3b}(s)}{s} R(s) ds.
\end{equation}
The~``timelike'' next--to--leading order kernel function entering this equation can be 
represented~as~\cite{Calmet1976}
\begin{equation}
\label{K3bRInt}
G_{3b}(s) = \int\limits_{0}^{1}\!
\frac{x^2 (1-x)}{x^2 + (1-x) s/\mmu^2}\,
\bar\Pi_{\ell}\biggl(\frac{x^2}{1-x}\,\mmu^{2}\biggr)
dx,
\end{equation}
where~$\bar\Pi_{\ell}(Q^{2})$~denotes the subtracted at zero leptonic vacuum 
polarization function (see,~e.g., Ref.~\cite{QED-AB1965})
\begin{align}
\label{LVP}
\bar\Pi_{\ell}(Q^2) & = 2 \!\int\limits_{0}^{1}\!
y(1-y)\ln\bigl[1+z_{\ell\,}y(1-y)\bigr] dy =
\nonumber \\ & =
-\frac{5}{9} + \frac{4}{3z_{\ell}} + \frac{2}{3}\biggl(1-\frac{2}{z_{\ell}}\biggr)
\sqrt{1+\frac{4}{z_{\ell}}}\,\arctanh\Biggl(\!\frac{1}{\sqrt{1+4/z_{\ell}}}\!\Biggr),
\quad\;\;
z_{\ell}=\frac{Q^2}{m^{2}_{\ell}} \ge 0
\end{align}
and $m_{\ell}$ stands for the mass of the corresponding lepton.

The specific form of the diagram displayed in the left plot of Fig.~\ref{Plot:Amu3bc} enables one 
to construct the explicit expression for the ``spacelike'' kernel function~$\KG{\Pi}{3b}(Q^2)$
straightaway. In~particular, since the diagram on hand 
factually constitutes an additional lepton loop insertion into the only internal photon line of 
the diagram shown in~Fig.~1, the function~$\KG{\Pi}{3b}(Q^2)$ is the product of the kernel 
function of the preceding order in the electromagnetic coupling~$\KG{\Pi}{2}(Q^2)$~(34) and the 
leptonic vacuum polarization function~$\bar\Pi_{\ell}(Q^2)$~(\ref{LVP}),~namely
\begin{equation}
\label{K3bPexpl}
\KG{\Pi}{3b}(Q^2) = \KG{\Pi}{2}(Q^2) \bar\Pi_{\ell}(Q^2).
\end{equation}
It~is worthwhile to note here that Eq.~(\ref{K3bPexpl}) has also been derived from 
the ``timelike'' expression~(\ref{K3bRInt}) by making use of the relevant
dispersion relations in Refs.~\cite{Chakraborty2018} and~[72].

\begin{figure}[t]
\centerline{\includegraphics[width=75mm,clip]{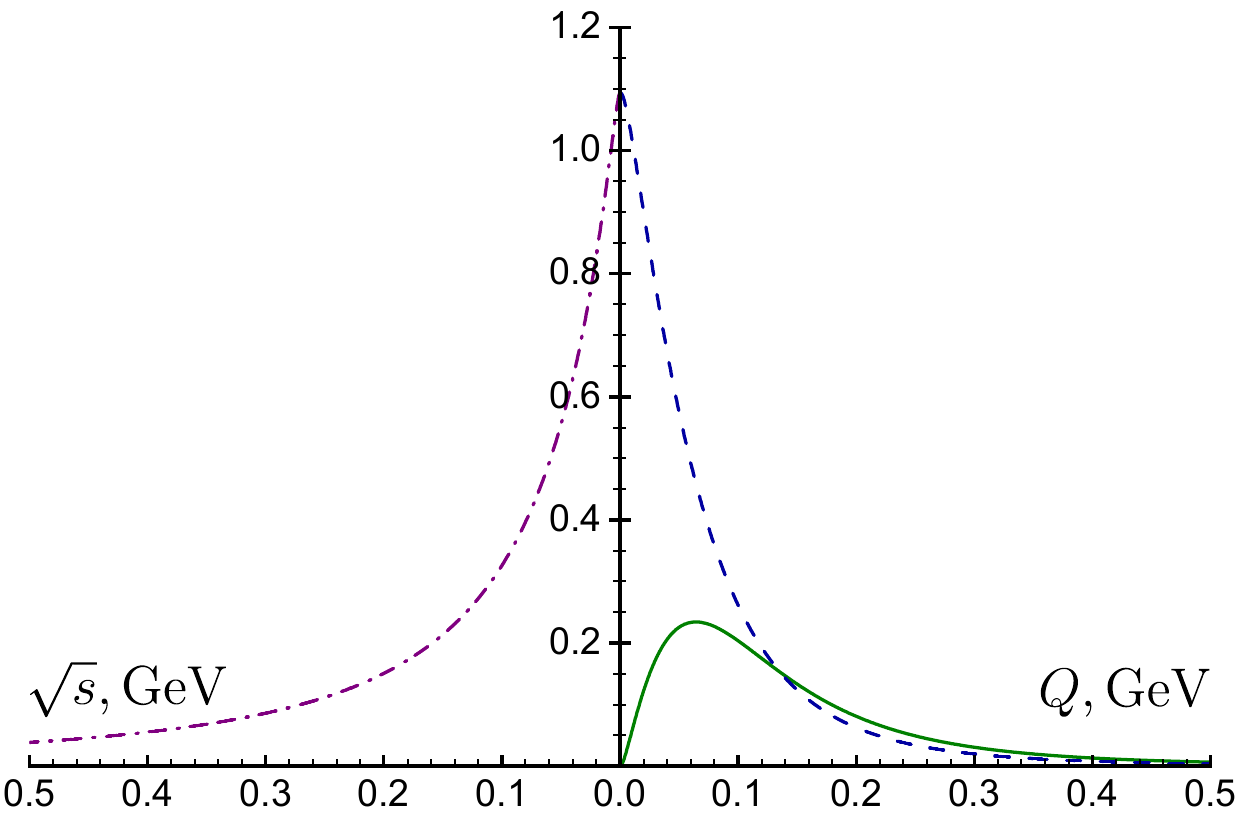}%
\hspace{10mm}%
\includegraphics[width=75mm,clip]{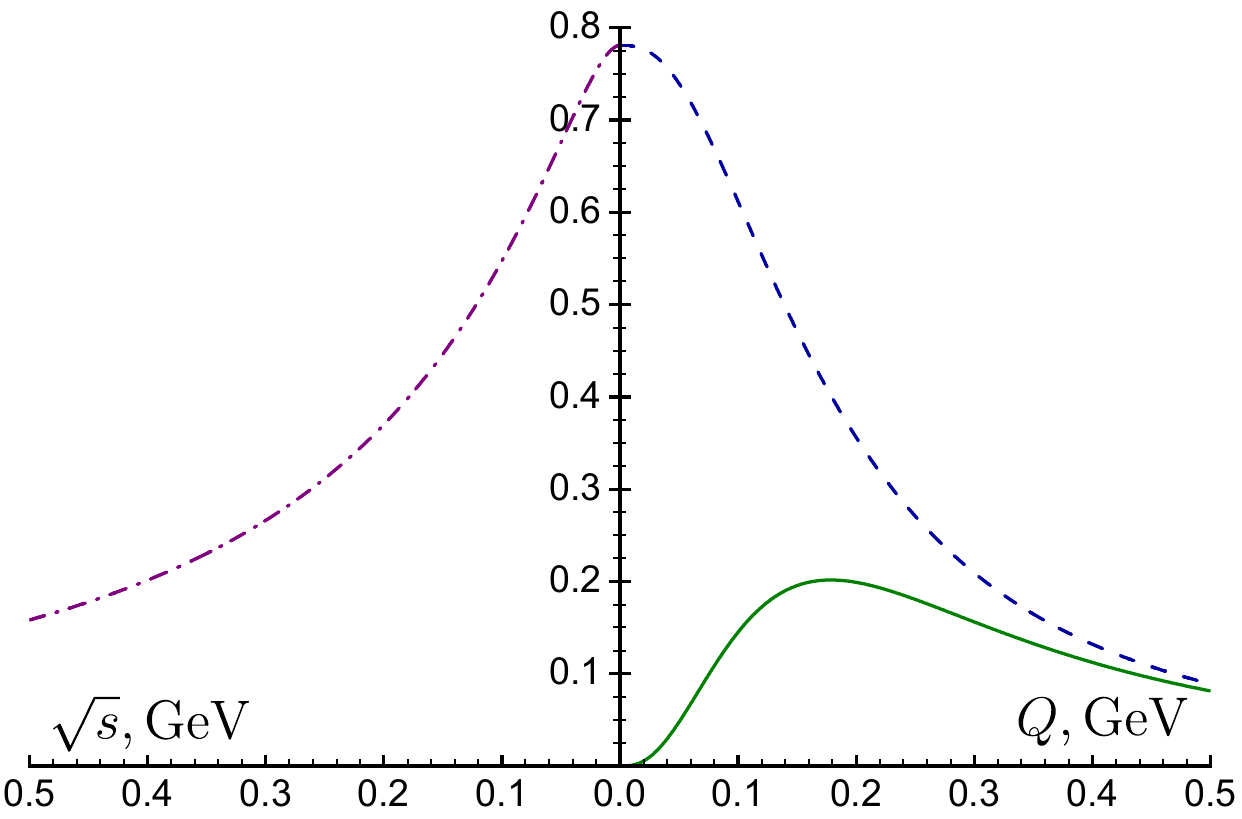}}
\caption{The kernel functions
$\zeta\KGt{\Pi}{3b}(\zeta)$ [Eq.~(\ref{K3bPexpl}), solid curve], 
$\zeta\KGt{D}{3b}(\zeta)$ [Eqs.~(\ref{KRelDP}),~(\ref{K3bPexpl}), 
dashed curve], and 
$\eta\KGt{R}{3b}(\eta)$ [Eq.~(\ref{K3bRInt}), dot--dashed curve]
in the 
spacelike [$Q^2 = -q^2 \ge 0$, \mbox{$\zeta = Q^2/(4\mmu^2)$}]
and timelike [$s = q^2 \ge 0$, $\eta = s/(4\mmu^2)$] domains.
The left plot corresponds to the electron loop insertion ($m_{\ell}=m_{e}$), whereas the right plot
corresponds to the $\tau$--lepton loop insertion ($m_{\ell}=m_{\tau}$, the displayed functions
are amplified by the factor of~$10^{4}$~here). The~notations of Eq.~(15) are~employed.}
\label{Plot:K3bPRD}
\end{figure}

The~plots of the ``spacelike'' [$\KG{\Pi}{3b}(Q^2)$, Eq.~(\ref{K3bPexpl})], 
[$\KG{D}{3b}(Q^2)$, calculated numerically by~making use 
of~Eqs.~(\ref{KRelDP}),~(\ref{K3bPexpl})] and 
``timelike''~[$\KG{R}{3b}(s)$, Eq.~(\ref{K3bRInt})] kernel 
functions are displayed in~Fig.~\ref{Plot:K3bPRD}. As~one can infer from this figure, 
in~the infrared limit the functions 
$Q^{2}\KG{D}{3b}(Q^2)$~[Eqs.~(\ref{KRelDP}),~(\ref{K3bPexpl})]
and $s\KG{R}{3b}(s)$~[Eq.~(\ref{K3bRInt})] assume the same 
value determined by~the relations~(31) and~(\ref{KRDlim}), specifically
\begin{align}
\label{Klim3b}
K_{0}^{(3b)} & =
\lim_{Q^2 \to 0_{+}} \frac{Q^2}{4\mmu^{2}}\, \KG{D}{3b}(Q^2) =
\lim_{s \to 0_{+}} \frac{s}{4\mmu^{2}}\, \KG{R}{3b}(s) =
\int\limits_{0}^{\infty}\!\! \KG{\Pi}{3b}(\xi)\, \frac{d \xi}{4\mmu^{2}}  = 
\nonumber \\ & =
-\frac{25}{36} -\frac{\ln\lambda}{3} +\lambda^{2}(4+3\ln\lambda) 
+\lambda^{4}\biggl[2\zeta_{2} -2\ln\lambda \ln\biggl(\frac{1}{\lambda}-\lambda\biggr) 
-\Li{2}(\lambda^{2}) \biggr]
+ \nonumber \\ &
+\frac{\lambda}{2}(1-5\lambda^{2})\biggl[ 3\zeta_{2} 
-\ln\lambda \ln\biggl(\frac{1-\lambda}{1+\lambda}\biggr) -\Li{2}(\lambda) +\Li{2}(-\lambda) \biggr],
\end{align}
that corresponds to the QED~contribution obtained in Refs.~\cite{Elend66, Massimo2004}. 
In~Eq.~(\ref{Klim3b}) $\lambda=m_{\ell}/\mmu$, $\zeta_{t}$~is the Riemann 
$\zeta$~function~(40), and~$\Li{2}(x)$ stands for the dilogarithm function~(38). 
For~the values of the lepton masses reported in Ref.~\cite{PDG22} the infrared limiting 
value~(\ref{Klim3b})~reads
\begin{equation}
\label{Klim3bNum}
K_{0}^{(3b)} \simeq
\begin{cases}
\displaystyle 1.094258,& \text{for~$m_{\ell}=m_{e}$},\\[1mm]
\displaystyle 0.780758 \times 10^{-4}, \quad& \text{for~$m_{\ell}=m_{\tau}$}.
\end{cases}
\end{equation}

\bigskip

As~for the third term on the right--hand side of Eq.~(\ref{Amu3Def}), it takes 
a particularly simple form in terms of the hadronic vacuum polarization 
function~$\bar\Pi(Q^2)$. Namely, following the very same lines as for 
the earlier discussed case of~$\KG{\Pi}{3b}(Q^2)$~(\ref{K3bPexpl}), one 
straightforwardly~gets
\begin{equation}
\label{Amu3cP}
\amu{3c} = \APF{3c}\!\!\int\limits_{0}^{\infty}\!
\KG{\Pi}{2}(Q^2) \Bigl[\bar\Pi(Q^2)\Bigr]^{\!2} \,\frac{d Q^2}{4\mmu^2},
\qquad
\APF{3c} = \frac{1}{9} \Bigl(\frac{\alpha}{\pi}\Bigr)^{\!3},
\end{equation}
with~$\KG{\Pi}{2}(Q^2)$ being the ``spacelike'' kernel function of the preceding 
order~(34). In~turn, Eq.~(\ref{Amu3cP}) can be expressed in terms of the 
function~$R(s)$~(4) by~making use of the dispersion relation~(2), namely
\begin{equation}
\label{Amu3cR}
\amu{3c} = \APF{3c} 
\int\limits_{s_{0}}^{\infty}\frac{d s_{1}}{s_{1}}\!
\int\limits_{s_{0}}^{\infty}\frac{d s_{2}}{s_{2}}\,
\KG{R}{3c}(s_{1}, s_{2}) R(s_{1}) R(s_{2}),
\end{equation}
where
\begin{equation}
\label{KR3cGen}
\KG{R}{3c}(s_{1}, s_{2}) = \int\limits_{0}^{\infty}\!
\frac{\KG{\Pi}{2}(Q^2)\,Q^{4}}{(s_{1}+Q^2)(s_{2}+Q^2)}\frac{d Q^2}{4\mmu^2},
\end{equation}
see Ref.~\cite{Calmet1976}. The~explicit form of the ``timelike'' kernel 
function~$\KG{R}{3c}(s_{1}, s_{2})$~(\ref{KR3cGen}) was given~in, e.g., Ref.~[59].
At~the same time, the contribution
to the muon anomalous magnetic moment~(\ref{Amu3cP}) can also be represented in terms of the 
Adler function~$D(Q^2)$~(6). Specifically, the relation, which expresses the hadronic 
vacuum polarization function~(1) in~terms of the Adler function~(6), 
reads~\cite{Pennington1977, Pennington1981, Pennington1984, 
Pivovarov1991}
\begin{equation}
\label{P_Disp2}
\Pi(-Q^{2}) - \Pi(-Q_{0}^{2}) = - \int\limits_{Q_{0}^{2}}^{Q^{2}} D(\xi)
\frac{d \xi}{\xi},
\end{equation}
where $Q^{2}=-q^{2}>0$ and~$Q_{0}^{2}=-q_{0}^{2}>0$ denote, respectively, 
the spacelike kinematic variable and the subtraction point. Then by~virtue 
of Eqs.~(3) and~(\ref{P_Disp2}) the contribution~(\ref{Amu3cP}) takes the
following form
\begin{equation}
\label{Amu3cD}
\amu{3c} = \APF{3c} \int\limits_{0}^{\infty} \frac{d Q^2}{4\mmu^2}
\,\KG{\Pi}{2}(Q^2)\!\! \left[\int\limits_{0}^{Q^2}\frac{d\xi}{\xi}D(\xi)\right]^{\!2}\!,
\end{equation}
with~$\xi = -p^{2} \ge 0$ being a spacelike kinematic variable.

\bigskip

The~obtained results can be employed in the studies of the hadronic 
vacuum polarization contributions to the muon anomalous magnetic 
moment~$a^{\text{HVP}}_{\mu}$. In~particular, the results presented
in this addendum, being complemented with those of Sect.~3.2 of 
Ref.~\cite{JPG49}, enable one to assess~$a^{\text{HVP}}_{\mu}$ 
at the next--to--leading order within spacelike methods, such as
lattice studies~[40, 41], MUonE project~[43--45], and others.

\end{document}